%% file: d0_top.tex
\documentclass[12pt]{article}
\usepackage{graphicx}

\newcommand\pubnumber{DPF2015-235}
\newcommand\pubdate{\today}

\def\nebraska{Department of Physics and Astronomy\\
University of Nebraska-Lincoln, Lincoln, NE 68588-0299}

\def\Title#1{\begin{center} {\Large #1 } \end{center}}
\def\Author#1{\begin{center}{ \sc #1} \end{center}}
\def\Address#1{\begin{center}{ \it #1} \end{center}}

\newcommand\pubblock{\rightline{\begin{tabular}{l} \pubnumber\\
         \pubdate  \end{tabular}}}
\newenvironment{Abstract}{\begin{quotation}  }{\end{quotation}}
\newenvironment{Presented}{\begin{quotation} \begin{center} 
             PRESENTED AT\end{center}\bigskip 
      \begin{center}\begin{large}}{\end{large}\end{center} \end{quotation}}
\def\Acknowledgments{\bigskip  \bigskip \begin{center} \begin{large}
             \bf ACKNOWLEDGMENTS \end{large}\end{center}}

\input econfmacros.tex

\begin{document}
\begin{titlepage}
\pubblock

\vfill
\Title{Recent Results on Top-Quark Physics at D0}
\vfill
\Author{Kenneth Bloom\\
for the D0 Collaboration}
\Address{\nebraska}
\vfill
\begin{Abstract}
  We present the most recent measurements on top-quark physics obtained
  with Tevatron $p\bar{p}$ collisions recorded by the D0 experiment at
  $\sqrt{s}=1.96$~TeV. The full Run II data set of 9.7 fb$^{−1}$ is
  analyzed. Both lepton+jets and dilepton channels of top-quark pair
  production are used to measure the differential and inclusive
  cross sections, the forward-backward asymmetries, the top-quark mass, the
  spin correlations, and the top-quark polarization.
\end{Abstract}
\vfill
\begin{Presented}
DPF 2015\\
The Meeting of the American Physical Society\\
Division of Particles and Fields\\
Ann Arbor, Michigan, August 4--8, 2015\\
\end{Presented}
\vfill
\end{titlepage}
\def\thefootnote{\fnsymbol{footnote}}
\setcounter{footnote}{0}

\section{Introduction}

In 2005, when the Tevatron was the only place to find a top quark, if you
you would have had to kill someone to give a talk with this title at the
DPF conference.  Ten years later, with huge samples of top quarks available from the
Large Hadron Collider, the author was asked if he could kindly present
these results from the D0 experiment, as he was going to the conference
anyway.  Why would anyone care about top quark measurements at D0 at this
point?  In fact D0 still has a number of interesting new results in
top-quark physics that take advantage of the unique proton-antiproton
initial state.  In addition, the highly-refined D0 analysis tools and
careful understanding of the veteran D0 detector allow for some very
sophisticated and precise measurements of top quark properties.  Thus, it
was a pleasure to give this presentation.

Many results were presented at the conference, but in the interest of time
and space, many of them will be tersely summarized here with references to
the relevant publications.  Only the newest results, especially those that
are new for Summer 2015, will be discussed in detail.

As a reminder to readers, the top quark is the heaviest fundamental
particle discovered so far, and the one with the largest Yukawa coupling to
the Higgs boson.  As a result, it is possible that it has some special role
to play in electroweak symmetry breaking.  It has a lifetime on the order
of a yoctosecond, which means that it undergoes a weak decay before it can
have any strong interactions.  This allows one to measure the properties of
a bare quark without messy QCD getting in the way.  Because the top quark
is so massive, it dominates loop diagrams that make quantum corrections to
the masses of the $W$, $Z$ and $H$ bosons.  Thus it plays a role in
standard-model self-consistency tests, and could provide clues to the
hierarchy problem and perhaps the stability of the electroweak vacuum.  All
of these make the study of the top quark compelling, even twenty years
after it was first discovered.

The D0 experiment operated at the Fermilab Tevatron through September 2011,
recording about 10~fb$^{-1}$ of $p\bar{p}$ collisions at
$\sqrt{s} = 1.96$~TeV.  The detector had a fairly small tracking volume and
magnetic field, but had an excellent calorimeter and extensive muon
coverage.  Table~\ref{tab:xsec} gives the predicted production cross
sections for strongly-produced $t\bar{t}$ and electroweakly-produced
single-top quarks at the Tevatron and at LHC Run~1.  While the cross
sections are generally much larger at the LHC, that is not the case for the
$s$-channel single-top production in the $tb$ mode, which is ``only'' a
factor of five larger compared to the Tevatron.  Also, at the Tevatron the
$q\bar{q}$ initial state provides about 85\% of the total cross section for
$t\bar{t}$ production, while it is only about 15\% at the LHC.  This gives
D0 unique access to the physics of that production mode.

\begin{table}[t]
\begin{center}
\begin{tabular}{crrrr}  \hline
& $t\bar{t}$ & $tb$ & $tqb$ & $tW$ \\
Tevatron & 7.08 & 1.04 & 2.08 & 0.30\\
LHC Run 1 & 234 & 5.55 & 87.2 & 22.2\\\hline
\end{tabular}
\caption{Cross sections, in picobarns, for the production of different
  final states including top quarks at the Tevatron ($p\bar{p}$, $\sqrt{s}
  = 1.96$~TeV) and the LHC Run~1 ($pp$, $\sqrt{s} = 8$~TeV), assuming a
  top-quark mass of 173~GeV~\cite{bib:kidonakis}.}
\label{tab:xsec}
\end{center}
\end{table}

\section{Production, Kinematics, Branching Ratios}

One process that the Tevatron experiments have unique access to is the
$s$-channel production of single top quarks; at the LHC, the backgrounds
(from $t\bar{t}$ production) are much more significant.  The rate for this
process is sufficiently small that results from the full datasets of both
Tevatron experiments, D0 and CDF, need to be combined to obtain a
measurement with sufficient statistical significance to be called an
observation of the process~\cite{bib:schannel}.  The cross section result,
$\sigma_{s \, \mathrm{channel}} = 1.29^{+0.26}_{-0.24}$~pb, with 6.3 standard
deviations significance.  This measurement then allows separate estimates
of the $s$-channel and $t$-channel cross sections, without any assumptions
of the value of $\sigma_s/\sigma_t$.  The results are consistent with the
standard model predictions, with no indication of any other contributing
process.  The two cross section values then leads to a measurement of
$V_{tb}$ that makes no assumptions on the number of quark generations,
unitarity, or $\sigma_s/\sigma_t$ (but does assume standard model top
decays, a pure $V-A$ interaction, and CP conservation).  The result is
$|V_{tb}| = 1.02^{+0.06}_{-0.05}$, or $|V_{tb}| \geq 0.92$ at 95\%
confidence level after applying a flat prior distribution for
$|V_{tb}| < 1$.

D0 has also performed a measurement of the $t\bar{t}$ cross section as a
function of various kinematic parameters such as $m_{t\bar{t}}$, $p_T(t)$
and $|y(t)|$ as a test of QCD~\cite{bib:differential}.  No signs of new
physics are observed.

\begin{figure}[htb]
\centering
\includegraphics[height=2.0in]{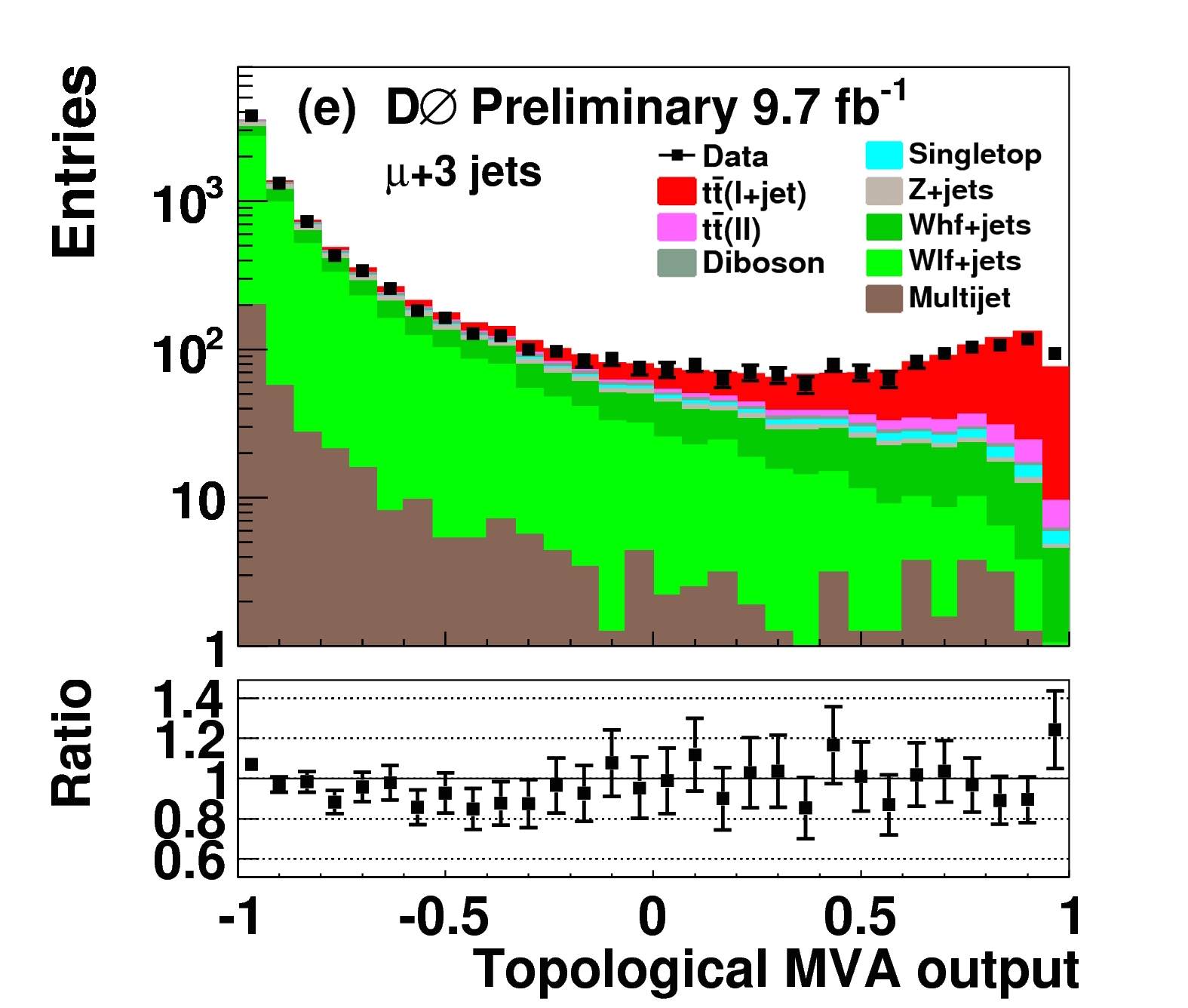}
\includegraphics[height=2.0in]{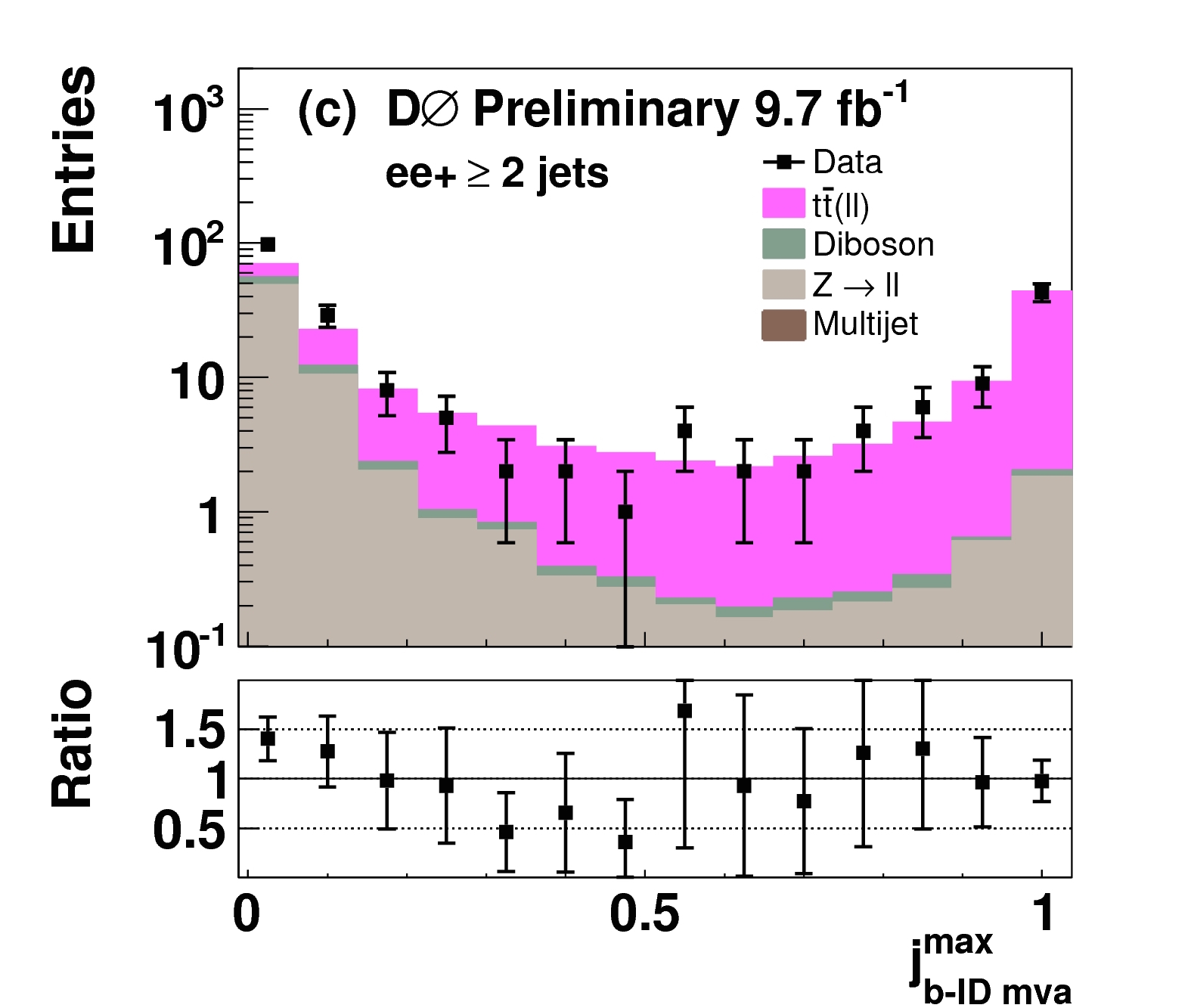}
\caption{Representative distributions from the inclusive $t\bar{t}$ cross
  section measurement.  Left: Output of the boosted decision tree with
  gradients for the $\mu$ plus three jets channel.  Right: $b$-tag
  discrimination variable for the leading jet in the $ee$ plus at least two
  jets channel.  In both cases the colored histograms indicate the
  contributions from different physics proccesses.}
\label{fig:xsecvars}
\end{figure}

A new preliminary measurement from D0, not yet published, gives a precise
measurement of the inclusive $t\bar{t}$ cross section that makes use of
both the dilepton and lepton-plus-jets channels~\cite{bib:ttxsec}. The
analysis makes heavy use of multivariate techniques, in which the numeric
values of many individual observables from an event are combined to form
one single quantity, and fits to distributions of those quantities from
each different final state are used to obtain the cross section.  The
lepton plus jets channel is broken into six subsamples based on lepton type
(electron or muon) and jet multiplicity (two, three or at least four jets).
Each subsample gets its own boosted decision tree with gradients using
about twenty kinematic variables, plus the output of a multivariate
algorithm used to identify $b$ jets.  The dilepton channel is simpler.  It
is broken into four subsamples ($e\mu$ plus one jet, $e\mu$ plus at least
two jets, $ee$ plus at least two jets and $\mu\mu$ plus at least two jets),
and the $b$-tag variable of the leading jet is the only one needed for the
fit.  Some representative distributions from the analysis are shown in
Figure~\ref{fig:xsecvars}.  The cross section is obtained from a
simultaneous log-likelihood fit template fit across all samples, using
systematic uncertainties as nuisance parameters.  The profiling of
systematic uncertainties reduces them by cross-calibration (for those that
are uncorrelated).  Careful attention is paid to correlations amongst
systematic uncertainties in the different subsamples.  The leading
systematic uncertainties are from signal modeling, especially
hadronization.

\begin{figure}[htb]
\centering
\includegraphics[height=2.5in]{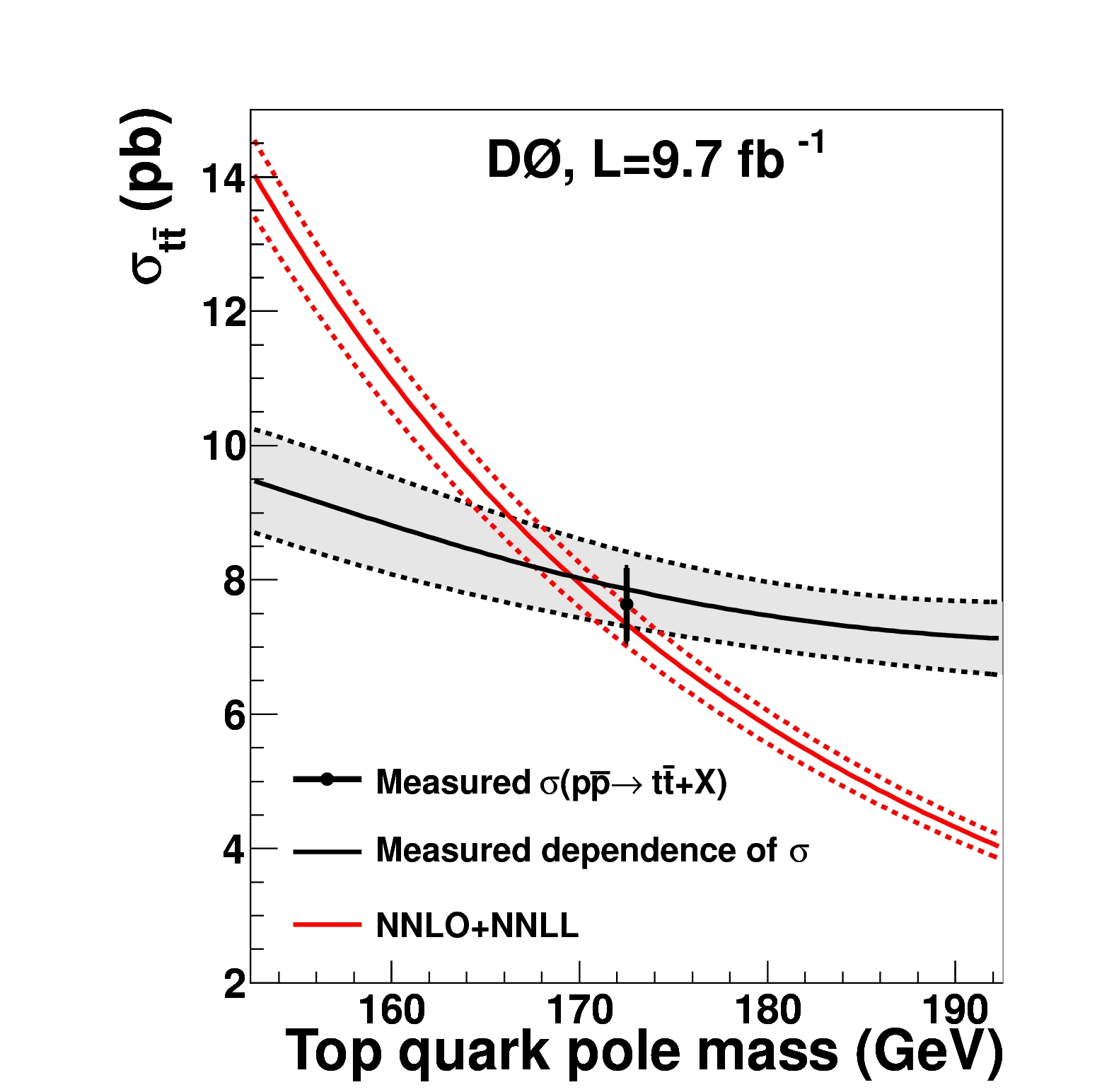}
\caption{The measured $t\bar{t}$ production cross section dependence on the
  top quark mass compared to the one provided by a next-to-next-to
  leading order perturbative QCD calculation.~\cite{bib:xsecvsmasscalc}}
\label{fig:polemass}
\end{figure}

The results for the analysis are given in Table~\ref{tab:xsecresults}.  The
cross sections are evaluated under the assumption of $m_t = 172.5$~GeV.
But one can also use the cross section measurement to estimate the
top-quark pole mass.  This approach to the mass measurement avoids
interpretation issues related to the definition of the quark mass.  The result is
illustrated in Figure~\ref{fig:polemass}, where the pole-mass dependence of
this measurement and of the cross section are displayed.  The result,
$m_t = 169.5^{+3.3}_{-3.4}$~GeV, is the most precise pole-mass
determination performed at the Tevatron.

\begin{table}[t]
\begin{center}
\begin{tabular}{lc}  \hline
Final state & Measured cross section (pb) \\\hline
Lepton + jets & 7.63 $\pm$ 0.14 (stat) $\pm$ 0.59 (syst)\\
Dilepton & 7.60 $\pm$ 0.34 (stat) $\pm$ 0.59 (syst)\\
Combined & 7.73 $\pm$ 0.13 (stat) $\pm$ 0.55 (syst)\\\hline
\end{tabular}
\caption{Results from the inclusive $t\bar{t}$ cross section measurement}
\label{tab:xsecresults}
\end{center}
\end{table}

\section{Top mass}

The most precise single measurement of the top-quark mass in the world (at
this moment) uses the matrix-element method in lepton plus jets
events~\cite{bib:ljetsmass}.  It provides a simultaneous measurement of the
top mass and a calibration factor for the jet-energy scale that is derived
by constraining the hadronic $W$ decay in the $t\bar{t}$ events to the
known value of the $W$ mass.  The result is $m_t = 174.98 \pm 0.58$ (stat)
$\pm 0.49$ (syst)~GeV, or $m_t = 174.98 \pm 0.76$~GeV.  The publications
referenced have very detailed information about the method itself and
various cross checks.

D0 has also recently performed a measurement of the top-quark mass using
dilepton events; a paper for publication was submitted shortly after the
conference~\cite{bib:dilmass}.  The measurement uses the neutrino weighting
method, in which an integration over the phase space of the two neutrino
rapidities is done on an event-by-event basis, and then a weight is
calculated based on the consistency of the expected missing energy for a
given mass value with the observed missing energy.  A key advance compared
to previous results using the dilepton final state is the transfer of the
jet-energy scale and its uncertainty from the lepton-plus-jets mass
measeasurement to this analysis.  This measurement then becomes statistics
limited, and thus much effort has gone into reducing statistical
uncertainties.  These efforts reduce the statistical component of the
uncertainty by 25\% compared to the previous form of the analysis.  Using
558 events, a value of $m_t = 173.32 \pm 1.36$ (stat) $\pm$ 0.85 (syst)~GeV,
or $m_t = 173.32 \pm 1.60$~GeV is obtained.  This result, with 0.92\%
precision, is consistent with the world average value, and the 0.49\%
systematic uncertainty is the smallest of any top-mass measurement in the
dilepton channel, making it competitive with results from the LHC.

\section{Asymmetries and Polarizations}

Due to interference terms that arise at next to leading order in QCD,
$t\bar{t}$ pairs produced from $q\bar{q}$ interactions have a
forward-backward asymmetry in the direction of the resulting quarks; the
$t$ tends to follow the direction of the $q$ and the $\bar{t}$ the
direction of the $\bar{q}$.  This asymmetry, defined as
\begin{equation}
A_{FB} = \frac{N(\Delta y > 0) - N(\Delta y < 0)}
{N(\Delta y > 0) + N(\Delta y < 0)},
\end{equation}
where $\Delta y = y_t - y_{\bar{t}}$, the rapidity difference between the
top quark and antiquark, is predicted to be about 10\%~\cite{bib:xsecvsmasscalc}.  The Tevatron has
unique access to this quantity as most of the $t\bar{t}$ pairs are produced
in $q\bar{q}$ interactions, which is not the case at the LHC.  The
$t\bar{t}$ forward-backward asymmetry has been a topic of great interest
for some years, as an anomalously large result could be an indicator of new
physics, and some early measurements of this quantity using the amount of
Tevatron data that was available at the time were in fact quite large.

D0 has now made several measurements of $A_{FB}$ using the full Tevatron
dataset.  A measurement with the lepton plus jets final state was completed
about a year ago~\cite{bib:ljetsafb}.  Compared to the previous measurement
in this final state, this one expanded the phase space used by including
events with only three jets in the final state, and did a careful
two-dimensional unfolding to parton level as a function of both $\Delta y$
and different kinematic variables.  The result,
$A_{FB} = (10.6 \pm 3.0)\%$, agrees with the standard model prediction
within uncertainties.  The asymmetry shows a stronger dependence on
$m_{t\bar{t}}$ and $|\Delta y|$ than predicted, but not in a statistically
significant way.  A measuerment of the forward-backward asymmetry for the
leptons in the same lepton plus jets events also yielded a result
consistent with the standard model prediction~\cite{bib:ljetsal}.

\begin{figure}[htb]
\centering
\includegraphics[height=1.4in]{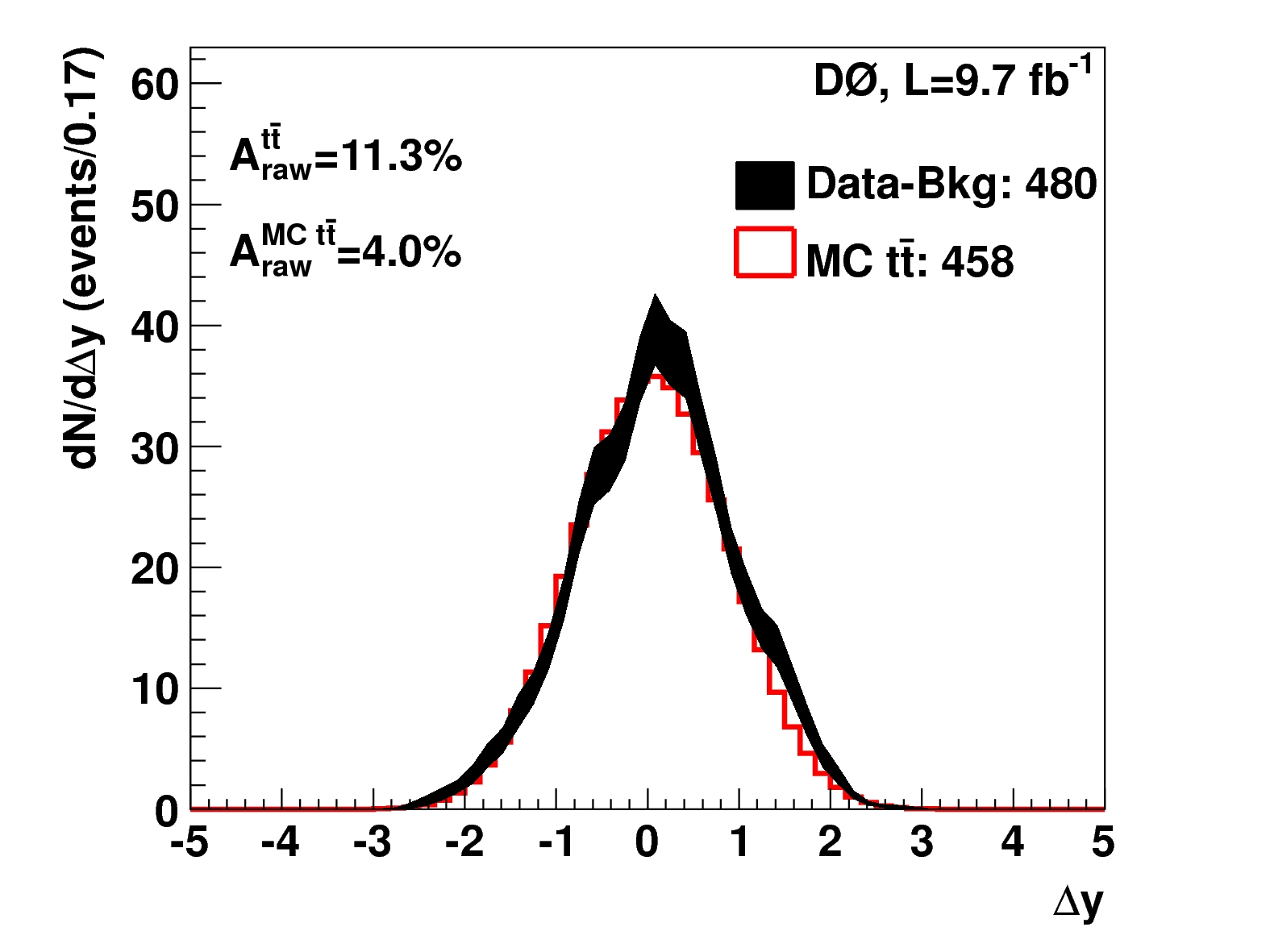}
\includegraphics[height=1.4in]{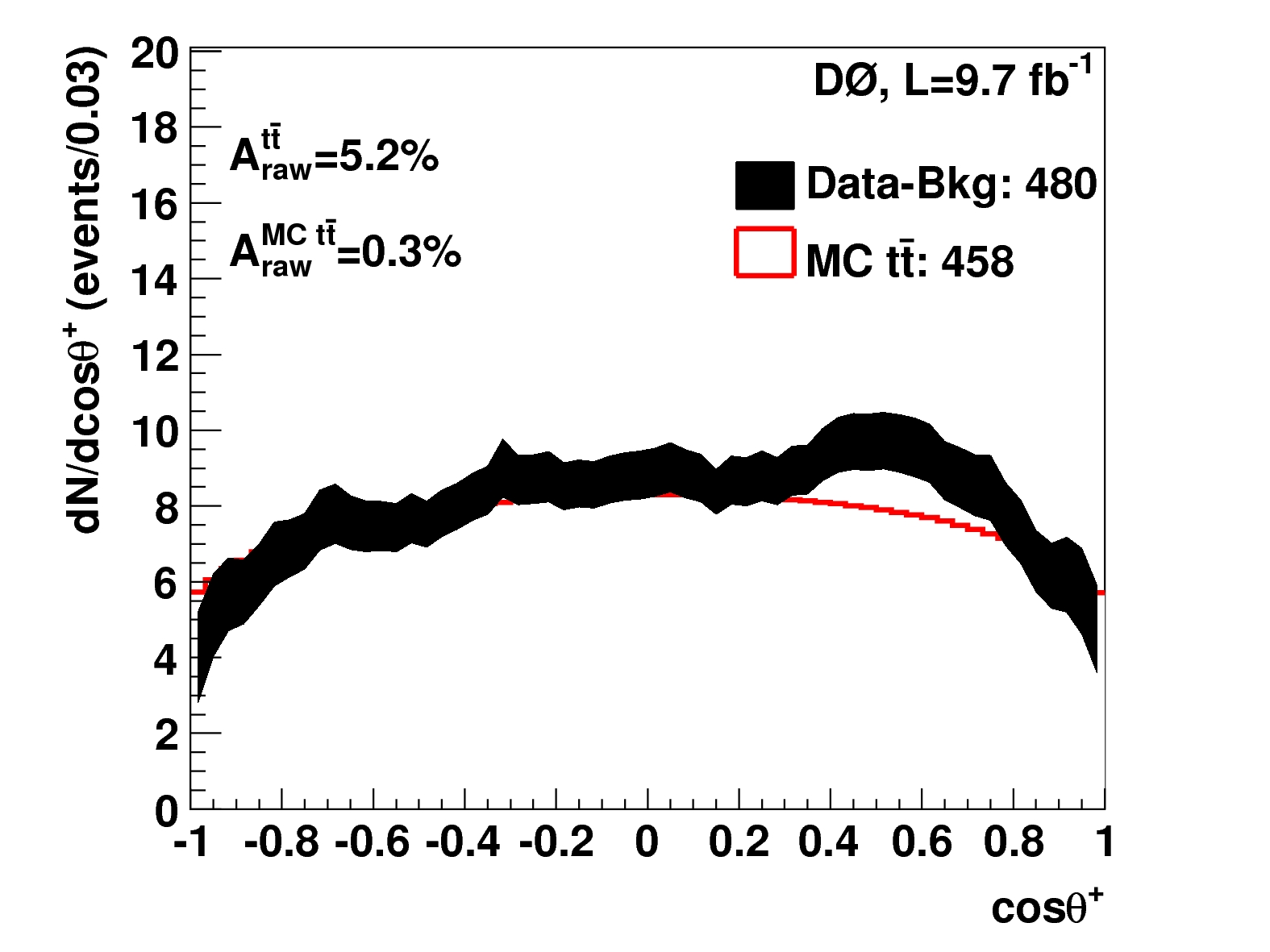}
\includegraphics[height=1.4in]{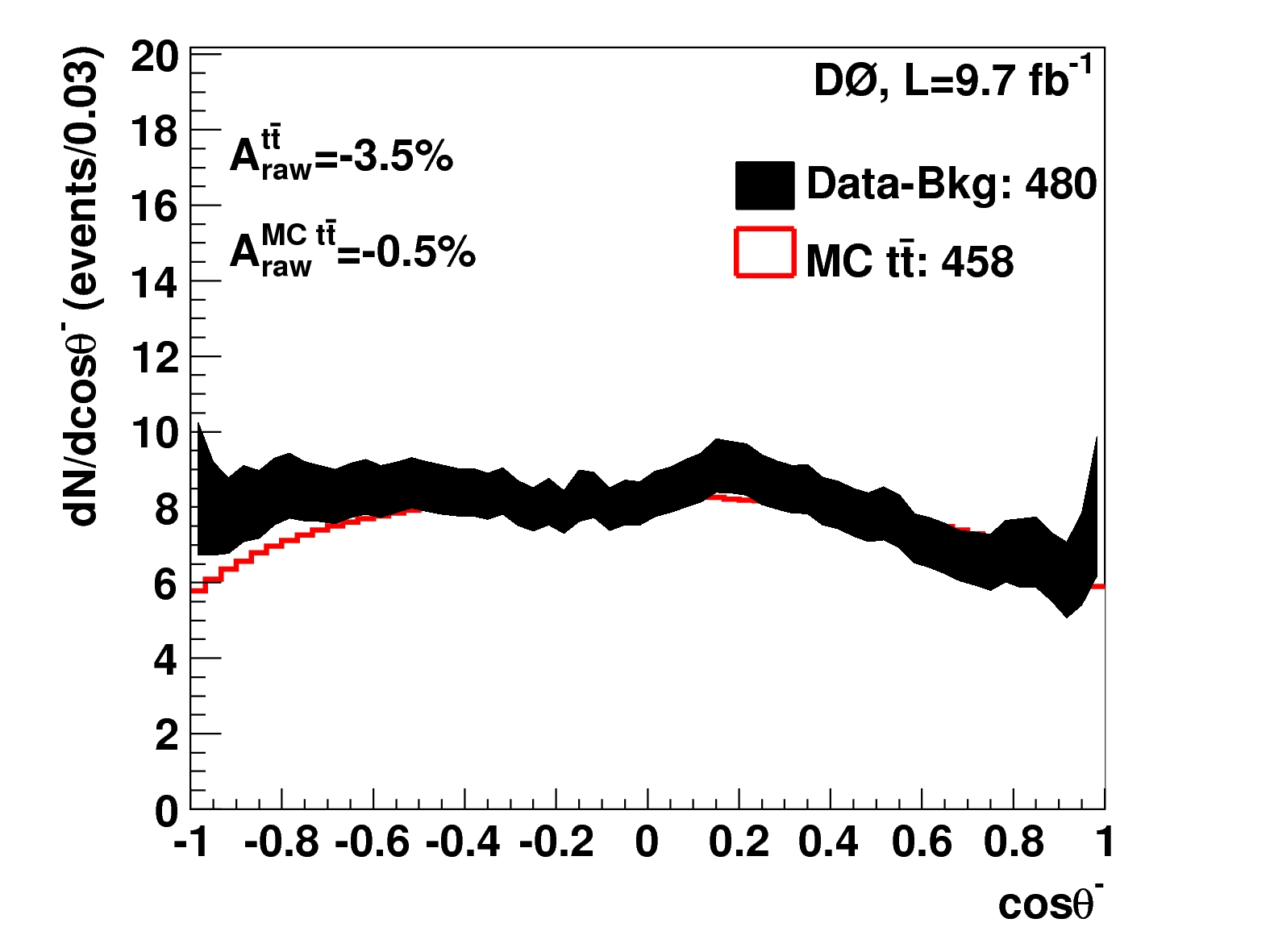}
\caption{Estimated distribution of $\Delta y_{t\bar{t}}$ (left),
  $\cos\theta^+$ (center) 
  and $\cos\theta^-$ (right) observables in dilepton events after
  subtracting the expected background contributions.  The
  background-subtracted data asymmetries and the Monte Carlo asymmetries
  extracted from these distributions are also reported.  These raw
  asymmetries need to be corrected for calibration effects to retrieve the
  parton-level asymmetries.}
\label{fig:dilkinematics}
\end{figure}

A new measurement of $A_{FB}$ using dilepton events was available for this
conference; the result has since been published~\cite{bib:dilafb}.  The
analysis actually measures the production asymmetry simultaneously with the
polarization of the top quark, making this the first measurement ever of
top polarization at the Tevatron.  The measurement is a novel application
of the matrix-element technique.  A full reconstruction of the event
kinematics is performed in a probabilistic fashion, and then a likelihood
per event for the most probable kinematic value is made for both the
asymmetry and the lepton decay angle with respect to the beam axis in the
$t\bar{t}$ rest frame.  The resulting distributions of the kinematic
quantities, summed over all events and corrected for background
contributions, is shown in Figure~\ref{fig:dilkinematics}.  After an
appropriate calibration of the method, the relevant quantities can be
extracted from the distributions. The systematic uncertainties are
dominated by those involved in modeling the $t\bar{t}$ signal, in
particular hadronization and showering, and also the calibration of the
method.

\begin{figure}[htb]
\centering
\includegraphics[height=2.5in]{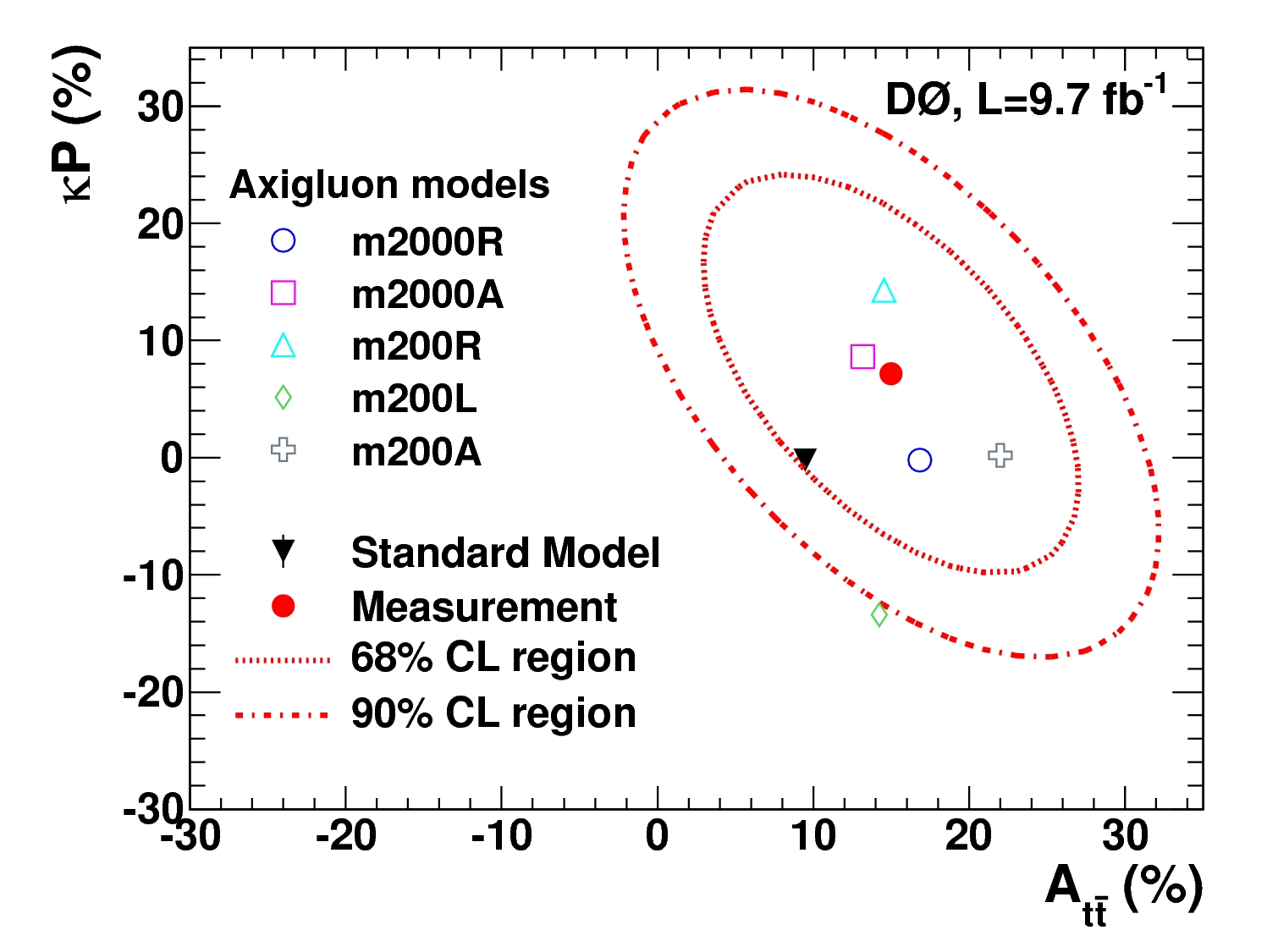}
\caption{Two dimensional visualization of the $A_{FB}$ and $\kappa P$
  measurements and comparison with benchmark axigluon models.  See
  Ref.~\cite{bib:dilafb} for more details about those models.}
\label{fig:afbpol}
\end{figure}

\begin{figure}[htb]
\centering
\includegraphics[height=2.5in]{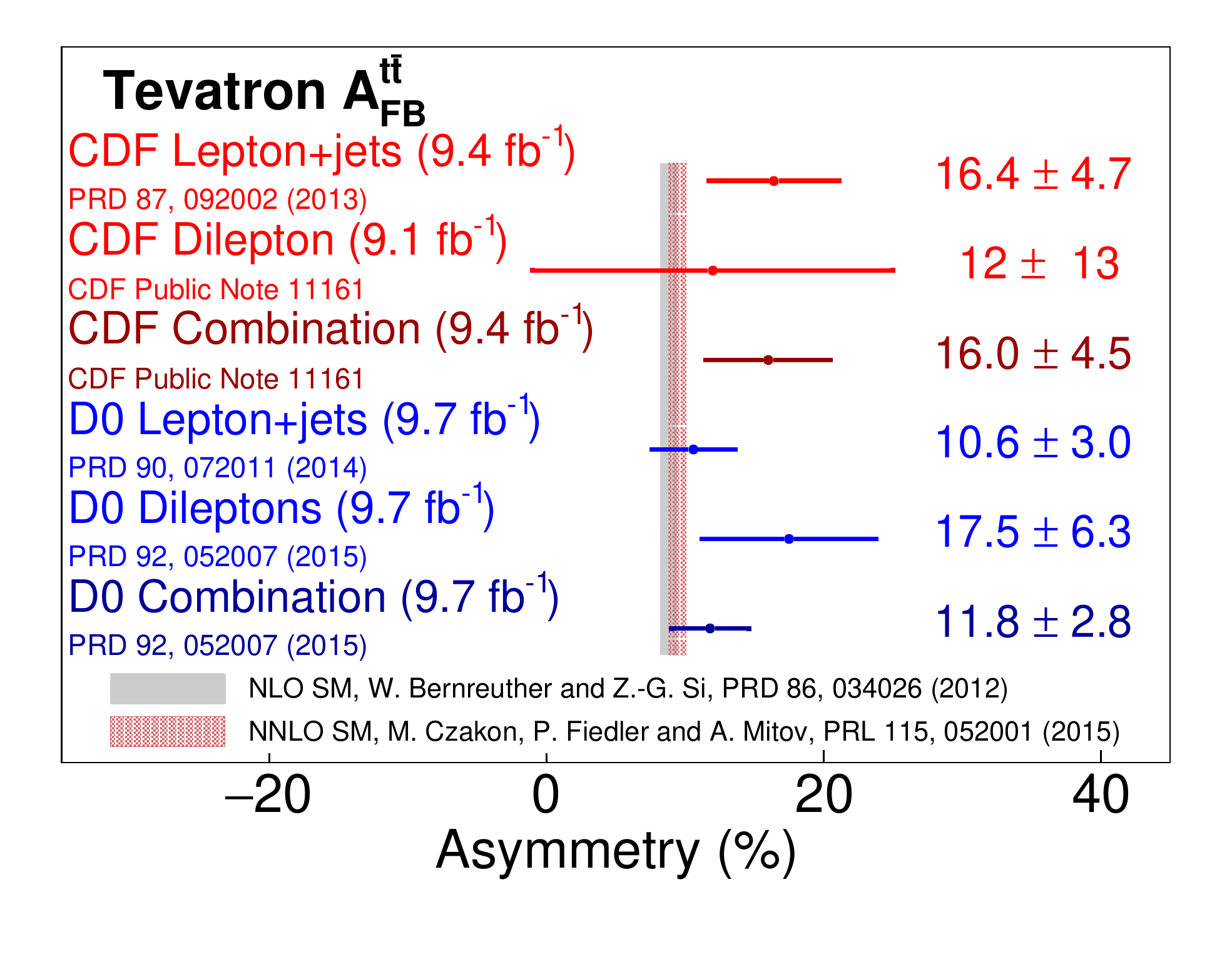}
\caption{Summary of $A_{FB}$ measurements at the Tevatron.}
\label{fig:afbsummary}
\end{figure}

The results of the measurement are shown in Figure~\ref{fig:afbpol}.
Without constraining either the asymmetry or the polarization, the results
are
\begin{eqnarray}
A_{FB} &=& (15.0 \pm 6.4 \pm 4.9)\%\\
\kappa P &=& (7.2 \pm 10.5 \pm 4.2)\%,
\end{eqnarray}
where the first uncertainty is statistical and the second is systematic,
and $\kappa \simeq 1.0$ is the spin analyzing power of
the lepton.  If one of the quantities is constrained to its standard-model
value, the result for the other quantity is
\begin{eqnarray}
A_{FB} &=& (17.5 \pm 5.6 \pm 3.1)\%\\
\kappa P &=& (11.3 \pm 9.1 \pm 1.9)\%.
\end{eqnarray}
The latter result for $A_{FB}$ is combined with that from the lepton plus
jets measurement to obtain the final D0 measurement of this quantity,
$A_{FB} = (11.8 \pm 2.5 \pm 1.3)\%$.  The complete set of $A_{FB}$
measurements from D0 and CDF are shown in Figure~\ref{fig:afbsummary}.  As
can be seen, there is a reasonable agreement between the results from the
two experiments, and between the experiments and predictions from theory.

\section{D0 top physics in the LHC era}
Even with the onslaught of data from the LHC, top physics at D0 is still
quite interesting.  The maturity of the experiment makes some very
sophisticated measurements possible, as the detector is very well-modeled
and the datasets are well-understood.  This allows for significant
creativity in data analyses, as exemplified by the fact that the
systematics-limited top mass measurements are quite competitive with those
from the LHC; one of them is currently the most precise single measurement
in the world.  In addition, the complementarity of the initial state
($p\bar{p}$ instead of $pp$) provides unique opportunities to search for
new physics.  The $t\bar{t}$ production asymmetry cannot be explored nearly
as well at the LHC, and the measurement of the $s$-channel single-top
production is very difficult there.  After twenty years of top physics, the
LHC experiments have much to learn from the Tevatron experience.  The last
few remaining D0 top-physics measurements should become available in the
coming months, completing a very successful physics program.

\Acknowledgments
I thank my D0 collaborators for all of their efforts to make these
measurements possible, and for giving me the opportunity to present them.
I am especially grateful to Andreas Jung, who provided significant
assistance in preparing the presentation.  My work on D0 is supported
by the National Science Foundation through award NSF-1306040.

\end{document}

%% file: econfmacros.tex
\def\beq{\begin{equation}}
\def\eeq#1{\label{#1}\end{equation}}
\def\eeqn{\end{equation}}

\def\beqa{\begin{eqnarray}}
\def\eeqa#1{\label{#1}\end{eqnarray}}
\def\eeqan{\end{eqnarray}}

\let\bar=\overbar

\def\Dslash{\not{\hbox{\kern-4pt $D$}}}
\def\dslash{\not{\hbox{\kern-2pt $\del$}}}

\def\msb{{\bar{\ssstyle M \kern -1pt S}}}




%% file: d0_top.bbl
\begin{thebibliography}{99}


\bibitem{bib:kidonakis} N.~Kidonakis, Phys. Part. Nucl. {\bf 45} 714
  (2014)  and references therein.
\bibitem{bib:schannel} CDF and D0 Collaborations, Phys. Rev. Lett. {\bf
    112}, 231803 (2014).
\bibitem{bib:differential} D0 Collaboration, Phys. Rev. {\bf D90}, 092006
  (2014).
\bibitem{bib:ttxsec} D0 Collaboration, D0 Note 6453-CONF (2015).
\bibitem{bib:xsecvsmasscalc} M.~Czakon and A.~Mitov, Computer Physics
  Communications {\bf 185}, 2930 (2014); P.~Barnreuther, M.~Czakon and
  A. Mitov, Phys. Rev. Lett. {\bf 109} 132001 (2012).
\bibitem{bib:ljetsmass} D0 Collaboration, Phys. Rev. Lett. {\bf 113},
  032002 (2014); D0 Collaboration, Phys. Rev. {\bf D91}, 112003 (2015); 
\bibitem{bib:dilmass} D0 Collaboration, arXiv:1508.03322 (2015).
\bibitem{bib:ljetsafb} D0 Collaboration, Phys. Rev. {\bf D90}, 072011 (2014). 
\bibitem{bib:ljetsal} D0 Collaboration, Phys. Rev. {\bf D90}, 072001 (2014). 
\bibitem{bib:dilafb} D0 Collaboration, Phys. Rev. {\bf D92}, 052007 (2015).

\end{thebibliography}
